\documentclass[prl,floatfix,twocolumn,superscriptaddress,showpacs,amsmath,amssymb]{revtex4-1}

\usepackage{graphicx}
\usepackage{epstopdf}
\usepackage{epsfig}
\usepackage{epsf}
\usepackage{url}
\usepackage[USenglish]{babel}
\usepackage{hyperref}
\def\bcen{\begin{center}}
\def\ecen{\end{center}}
\allowdisplaybreaks
\renewcommand\[{\begin{equation}}
\renewcommand\]{\end{equation}}
\usepackage{verbatim}
\usepackage{natbib}
\usepackage{amsmath, nccmath}
\usepackage[export]{adjustbox}
\usepackage{dcolumn}
\usepackage{bm}
\usepackage{bbm}
\usepackage{lipsum}
\usepackage{tikz-feynman}

\begin{document}
\title{Correlated electronic structure of La$_3$Ni$_2$O$_7$ under pressure}
\author{Viktor Christiansson}
\affiliation{Department of Physics, University of Fribourg, 1700 Fribourg, Switzerland}
\author{Francesco Petocchi}
\affiliation{Department of Quantum Matter Physics, University of Geneva, 1211 Geneva 4, Switzerland}
\author{Philipp Werner}
\affiliation{Department of Physics, University of Fribourg, 1700 Fribourg, Switzerland}

\begin{abstract}
Recently, superconductivity with a $T_c$ up to 78~K has been reported in bulk samples of the bilayer nickelate La$_3$Ni$_2$O$_7$ at pressures above 14 GPa. 
Important theoretical tasks are the formulation of relevant low-energy models 
and the clarification of the normal state properties. 
Here, we study the correlated electronic structure of the high-pressure phase in a four-orbital low-energy subspace using different many-body 
approaches: $GW$, dynamical mean field theory (DMFT), extended DMFT (EDMFT) and $GW$+EDMFT, with realistic frequency-dependent interaction parameters. 
The nonlocal correlation and screening effects captured by $GW$+EDMFT result in
an instability towards the formation of charge stripes, with the $3d_{z^2}$ as the main active orbital.
We also comment on the potential relevance of the rare-earth self-doping pocket, 
since hole doping suppresses the ordering tendency.
\end{abstract}

\maketitle
%
%
%
%
{\it Introduction}
Nickelates have been theoretically proposed more than 20 years ago as an interesting material platform to search for high-$T_c$ superconductivity \cite{Anisimov1999}, but they have moved into the spotlight of the condensed matter community only after the recent experimental discovery of superconductivity in thin films of Nd$_{1-x}$Sr$_{x}$NiO$_2$ and  Pr$_{1-x}$Sr$_{x}$NiO$_2$ with $x\approx 0.2$ \cite{Li2019,Zeng2020,Osada2020a}.
Apart from these hole-doped infinite-layered systems, superconductivity has also been demonstrated in the quintuple layer compound Nd$_6$Ni$_5$O$_{12}$ \cite{Pan2022}, which realizes a favorable Ni-$d$ filling for superconductivity, even without chemical doping \cite{Worm2022}. The highest $T_c$ values measured in these compounds so far are, however, substantially lower than in cuprates (maximum $T_c$ of $31$~K realized under pressure \cite{Wang2022}), and superconductivity has not been observed in bulk crystals. The reported bulk superconductivity with a $T_c$ near 80~K in the bilayered perovskite nickelate La$_3$Ni$_2$O$_7$ under pressure by Sun {\it et al.} \cite{Sun2023}, and shortly afterwards by Zhang and co-workers \cite{Zhang2023b}, thus appears to be a major breakthrough, which will stimulate the research on high-temperature superconductivity \cite{Luo2023,Zhang2023,Lechermann2023}. 

An important theoretical challenge is the formulation of a minimal model which captures the relevant low-energy physics. In the case of cuprates, many researchers believe that a single-band Hubbard model, effectively representing the physics of the $d_{x^2-y^2}$ and oxygen states, provides an adequate starting point \cite{Dagotto1994}. Indeed, qualitatively correct phase diagrams have been obtained with many-body treatments of this model \cite{Lichtenstein2000}. In the case of the infinite-layer nickelates, the situation is less clear. It has been pointed out early on that due to the larger energy separation between the $d$ and $p$ bands, the low-energy physics of the nickelates differs from that of hole-doped charge-transfer insulators like the cuprates \cite{Lee2004}. The important role of multi-orbital effects, involving especially the $d_{z^2}$ orbital in hole-doped nickelates, has also been emphasized \cite{Werner2020,Lechermann2020}. On the other hand, based on the close similarity between the experimental $T_c$ dome and the results from single-band calculations, it has been argued that the $d_{x^2-y^2}$ orbital plays the dominant role in nickelates, and that the single-band Hubbard model represents an adequate starting point \cite{Held2022}.  

\begin{figure}[b]
\begin{centering}
\includegraphics[width=0.9\columnwidth]{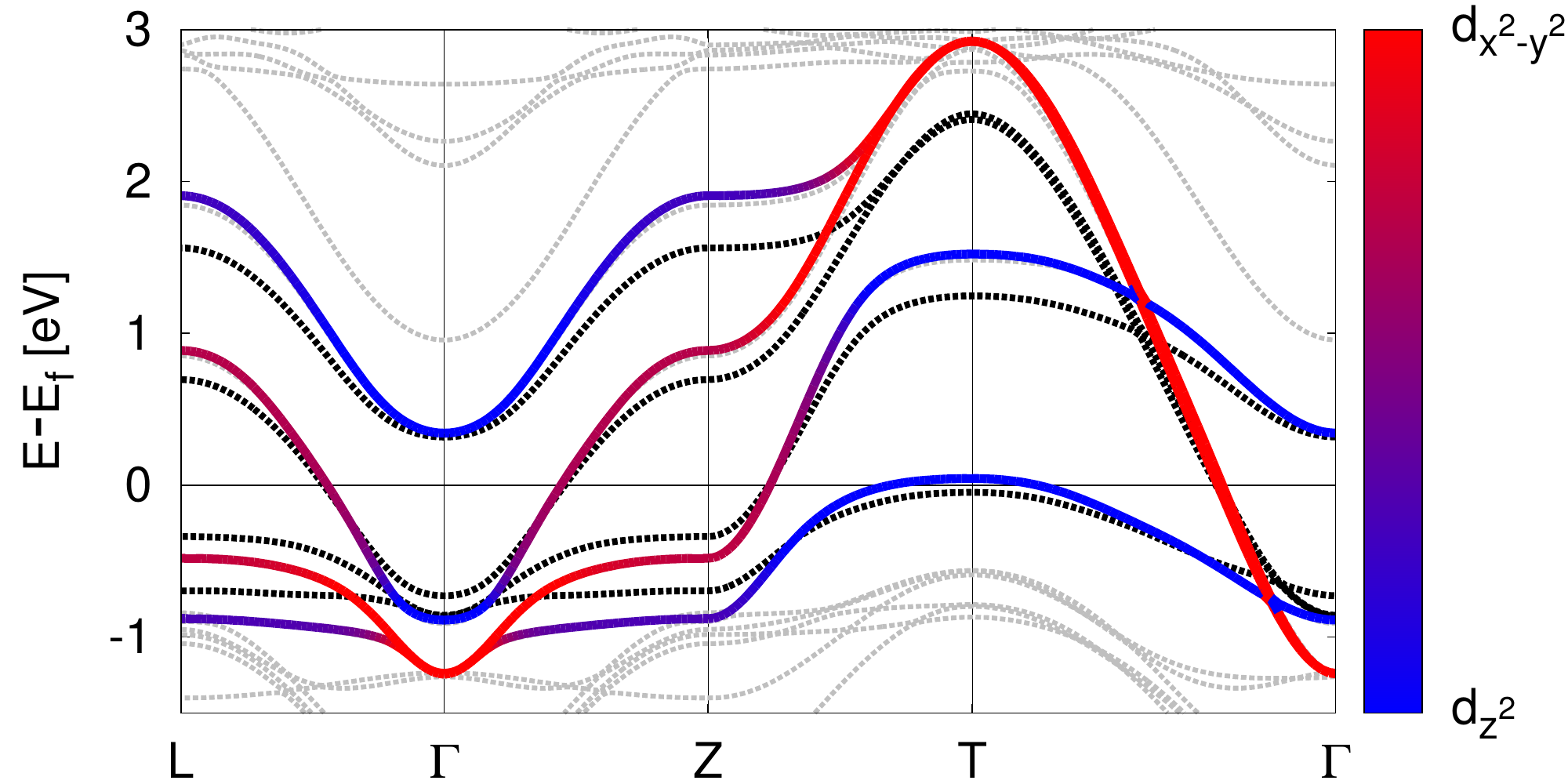}
\end{centering}
\caption{The model band structure (solid lines) plotted on top of the DFT band structure (gray dashed lines). The color-bar shows the orbital character of the model bands. We also indicate the $G^0W^0$ quasi-particle band structure by the black dashed line, showing a narrowing of the model bands and the disappearance of the hole pocket at the $T$-momentum.
\label{fig:BS}}
\end{figure}

Density functional theory (DFT) calculations for the high-pressure $Fmmm$ phase of the La$_3$Ni$_2$O$_7$ bilayer system suggest a low-energy model involving two $d_{x^2-y^2}$ and two $d_{z^2}$ Ni orbitals \cite{Sun2023}, where the latter form bonding and antibonding states due to the strong inter-layer hopping through the apical oxygen. There is a total of three electrons occupying the $d_{x^2-y^2}$ and the bonding states, which at the DFT level results in three partially filled metallic bands.
An important question is how this picture is modified if the electronic correlations are treated with advanced many-body techniques. Here, we calculate the dynamically screened interaction parameters for the four-orbital low-energy model from first principles \cite{Aryasetiawan2004}, and use them to compute the electronic structure with the $GW$ approximation \cite{Hedin1965}, dynamical mean field theory (DMFT) \cite{Georges1996}, extended DMFT (EDMFT) \cite{Sun2002} and $GW$+EDMFT \cite{Biermann2003,Boehnke2016,Nilsson2017}.

{\it DFT calculation and cRPA interactions}
We start from a DFT calculation \cite{Hohenberg1964,Kohn1965} of the high-pressure $Fmmm$ phase of bilayer La$_3$Ni$_2$O$_7$, which contains a single formula unit in the primitive cell with one Ni atom in each of the two layers. We perform calculations for the experimentally reported lattice parameters \cite{Sun2023}, at the three pressures $P=20.9$,  29.5, and 41.2 GPa, covering the range where superconductivity has been observed. The atomic positions are fully relaxed. 
Compared to its infinite-layered counterparts, the rare-earth self-doping pocket at the $\Gamma$-point in the DFT band structure is lifted above the Fermi level.  
It is important to note, however, that if we fix the atomic positions at the experimentally reported values \cite{Sun2023}, this band is lowered by 1 eV, leading again to a self-doping situation, see Supplemental Material (SM).
We will comment on the potential relevance of this pocket in the discussion of the $GW$+EDMFT results. 

Using the relaxed structure, we define a four orbital low-energy model of $d_{x^2-y^2}$- and $d_{z^2}$-like Wannier functions \cite{Marzari1997,Mostofi2008} on the two Ni sites. In Fig.~\ref{fig:BS} we show the resulting Wannier band structure (solid lines) overlayed on the DFT one (gray dashed lines).
The full band structure is then downfolded to this low-energy subspace using the multitier approach of Refs.~\cite{Boehnke2016,Nilsson2017}, 
with a constrained random-phase approximation (cRPA) \cite{Aryasetiawan2004} calculation to obtain the momentum- and frequency-dependent effective bare interaction parameters, and in the case of the $GW$-based methods a one-shot $G_0W_0$ calculation \cite{Hedin1965} to compute the effective bare propagators $G^0_{\bf k}(\omega)$ of the model. 
In the (E)DMFT calculations, unless otherwise noted, we solve separate impurity problems for each lattice site (see SM and Refs.~\cite{Petocchi2020b,Christiansson2023} for additional details).
In the case of $GW$ and $GW$+EDMFT the approach is free from double-countings and adjustable parameters (apart from the choice of the low-energy subspace), while in DMFT and EDMFT we use the fully localized limit double counting prescription \cite{Anisimov1993,Czyzyk1994,Werner2016}.

\begin{figure}
\begin{centering}
\includegraphics[width=\columnwidth]{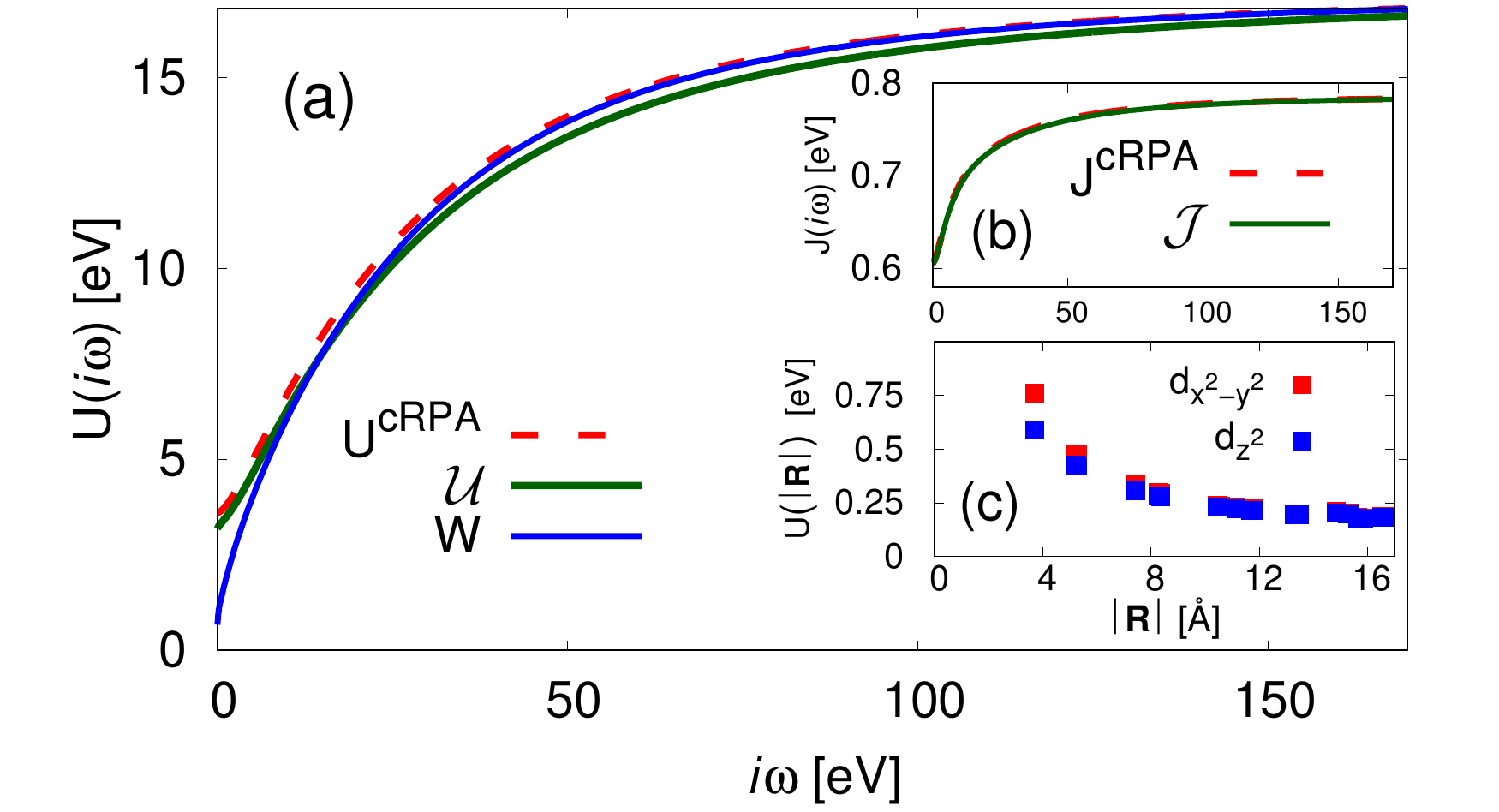}
\end{centering}
\caption{
(a) Frequency dependence of the bare local interactions for the $d_{z^2}$-like orbital, calculated within cRPA ($U$) and $GW$+EDMFT ($\mathcal{U}$), as well as the fully screened interaction $W$. The insets show (b) the cRPA $J$ and the $GW$+EDMFT effective Hund's couplings $\mathcal{J}$. (c) The non-local interaction $U(|{\bf R}|)$ within a single layer at different lattice spacings in \AA~for the $d_{x^2-y^2}$- and $d_{z^2}$-like orbitals.
\label{fig:interactions}}
\end{figure}

The effective interaction parameters of a low-energy model typically exhibit a large frequency dependence, especially for the density-density terms, while the Hund coupling parameter $J$ is less screened. In Fig.~\ref{fig:interactions} we show the cRPA result for the local (${\bf R}=0$) components of the $d_{z^2}$-like orbital, and compare it to the self-consistently calculated effective bare interaction $\mathcal{U}$ and fully screened interaction $W$ from the $GW$+EDMFT calculations. 
The static $(\omega=0)$ values of the local interactions are 
\begin{equation*}
U_{\mathrm{cRPA}}{=}\begin{pmatrix}
 d_{x^2-y^2} & \quad d_{z^2} \\
\hline
{3.79} & 2.39   \\
\cline{0-0} \multicolumn{1}{c|}{0.61} &  3.58
\end{pmatrix},\quad
\mathcal{U}{=}\begin{pmatrix}
 d_{x^2-y^2} & \quad d_{z^2} \\
\hline
{3.46} & 2.04   \\
\cline{0-0}  \multicolumn{1}{c|}{0.61} & 3.18
\end{pmatrix},
\label{eqn:UcrPA} 
\end{equation*}
where the upper triangular part lists the density-density interactions and the lower left entry the Hund's coupling. 
Since $\mathcal{U}$ takes into account non-local screening within the model subspace, the density-density terms are reduced compared to cRPA. 
The non-local components of the cRPA interaction are comparatively large, with an in-plane nearest-neighbor interaction of $\sim0.75$ eV for the $d_{x^2-y^2}$-like orbital, indicating a need for considering non-local screening effects. The long-ranged ($\sim \frac{1}{r}$) tail of the bare interactions is demonstrated in Fig.~\ref{fig:interactions}(c) for translations within the bilayer.

The static interaction parameters remain almost unaffected by hydrostatic pressure, with changes of less than  $0.05$ eV, in contrast to the infinite-layered system where we found non-trivial systematic changes as a function of (uniaxial) pressure \cite{Christiansson2023}. 
More detailed information on the local and non-local interactions and their pressure dependence is provided in the SM.

{\it GW results}
As a first test of the (nonlocal) screening and correlation effects, we perform self-consistent $GW$ calculations for our model, taking into account the full band structure through the downfolded interactions and propagators. The quasi-particle band structure corresponding to the initial $G^0_{\bf k}(\omega)$ is shown by the black dashed lines in Fig.~\ref{fig:BS}. The main effect of self-consistency is a slight renormalization of the bands and a transfer of charge between the Ni orbitals, bringing them closer to quarter and half-filling, respectively, as indicated in Tab.~\ref{Tab:Occupations}. Overall, the physical picture remains similar to DFT, with a filled bonding band of $d_{z^2}$ character and two approximately quarter-filled bands of $d_{x^2-y^2}$ character \cite{Sun2023}. The fact that the low-energy model has a bandwidth of $\sim 4$ eV, comparable to the on-site interactions, however suggests that an accurate treatment of correlation effects requires calculations beyond DFT or $GW$.

\begin{table}[t]
\caption{\label{Tab:Occupations} 
Orbital occupation per spin $n^i$ at inverse temperature $\beta=50$ eV$^{-1}$ ($\sim 230$ K) and pressure $P=29.5$ GPa for the Ni$_i$ $3d_{x^2-y^2}$- and $3d_{z^2}$-like orbitals in the two layers $i$ of the primitive unit cell. The total filling is $n=3$. $GW$+EDMFT results are shown for calculations with ordering (period-2 oscillations) and for suppressed order (SO), as discussed in the text.
The calculations for the supercell, performed at $\beta=10$ eV$^{-1}$, also result in period-2 oscillations. In this case we show the average occupations for one of the Ni-dimers.
}

\setlength{\tabcolsep}{5.4pt} 
\renewcommand{\arraystretch}{1.6} 
\centering
\begin{tabular}{|c||c c||c c|}
\hline
Method & $n^{1}_{x^2-y^2}$ & $n^{1}_{z^2}$ & $n^{2}_{x^2-y^2}$ & $n^{2}_{z^2}$ \\
\hline
\hline
DFT & 0.31 & 0.44 & 0.31 & 0.44  \\
\hline
sc$GW$ & 0.23 & 0.52 & 0.23 & 0.52  \\
\hline
DFT+DMFT $U^\textrm{cRPA}(0)$ & 0.32 & 0.43 & 0.32 & 0.43  \\
DFT+DMFT $U^\textrm{cRPA}(\omega)$ & 0.37 & 0.38 & 0.37 & 0.38  \\
\hline
EDMFT &  0.50  & 0.51 & 0.32 & 0.17  \\
\hline
$GW$+EDMFT & 0.96 & 0.47 & 0.04  & 0.03   \\
Supercell $GW$+EDMFT & 0.90 & 0.53 & 0.04 & 0.03 \\
$GW$+EDMFT (SO) & 0.08 & 0.67 & 0.08 & 0.67   \\
\hline
\end{tabular}
\end{table}

\begin{figure}[b]
\begin{centering}
\includegraphics[width=\columnwidth]{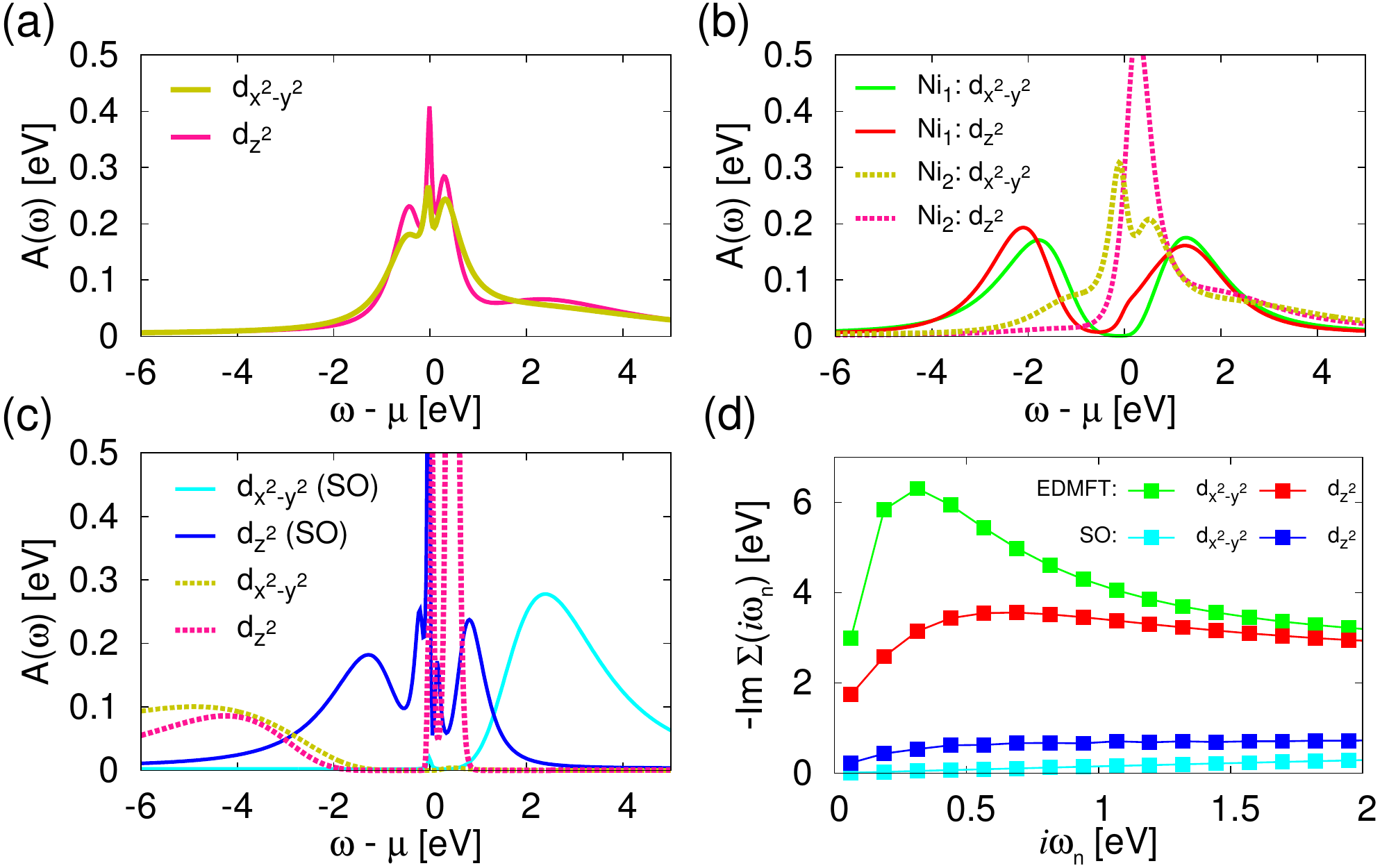}
\end{centering}
\vspace*{-0.5cm}
\caption{
Local spectral functions $A(\omega)$ for (a) DMFT+$U(\omega)$, (b) EDMFT, and (c) $GW$+EDMFT for the period-2 solution, and the solution with suppressed order (SO). 
(d) Imaginary part of the EDMFT self-energy (corresponding to the spectra in (b)) and $GW$+EDMFT result for suppressed order (corresponding to the spectra in (c)).
}
\label{fig:spectra_ordering}
\end{figure}

{\it DMFT and EDMFT results}
DMFT and EDMFT calculations provide a better description of local correlations by calculating a local self-energy (and local polarization function in the case of EDMFT) through a mapping of the lattice system to self-consistently determined impurity models \cite{Georges1996,Sun2002,Werner2016}. We treat two separate impurity problems for the Ni sites in the two layers and do not use any mixing in the self-consistency loop, in order to detect potential ordering instabilities \cite{Chan2009}. 
Our calculations down to inverse temperature $\beta=50$ eV$^{-1}$ ($\sim 230$~K) show that for the static cRPA interactions, the orbital occupations within DFT+DMFT are very similar to the DFT results and  previously reported DFT$+U$ results \cite{Sun2023}, and that there is no ordering instability. 
We have checked that varying the static $U$ values in a physically reasonable range between 3 and 5 eV does not change this picture (see SM). 
In Tab.~\ref{Tab:Occupations} we report the occupations obtained with both the static and frequency dependent local cRPA interactions.  
Taking into account the frequency dependence of the cRPA interactions leads to a transfer of charge to the $d_{x^2-y^2}$ orbital, resulting in a roughly equal filling in both orbitals.
The latter result is qualitatively similar to recent DMFT calculations in Ref.~\cite{Lechermann2023}, which however produced half-filled $e_g$ orbitals. We believe that the discrepancy is due to the larger model subspace and different interactions parameters used in Ref.~\cite{Lechermann2023}, which should also explain why our local DMFT spectra, shown in Fig.~\ref{fig:spectra_ordering}(a), are more metallic.

In the EDMFT calculations we take into account the local and nonlocal frequency dependent cRPA interactions, and treat dynamical screening effects in the low-energy subspace in the form of a 
local polarization. The nonlocal interactions lead to qualitative changes in the physics, compared to the previous methods. 
In particular, we now observe a breaking of the symmetry between the two Ni sites in the unit cell, corresponding to an emptying of one layer, and resulting in close to half-filled orbitals in the other layer.
(Such a charge reshuffling between layers would likely result in a lattice response, which is not captured in our model.)
The almost half-filled layer exhibits long-ranged antiferromagnetic (AFM) magnetic order, as indicated by oscillations between solutions with opposite spin polarizations in the selfconsistency loop. 
The local spectral functions and self-energies shown in Fig.~\ref{fig:spectra_ordering}(b,d) reveal that the half-filled layer is nearly Mott insulating \footnote{Without particle-hole symmetry, even in the Mott state, $-\text{Im}\Sigma(i\omega_n)$ does not diverge as $\omega_n\rightarrow 0$.}, which appears to be the driving force behind the ordering instability.

{\it $GW$+EDMFT results}
Previous model and {\it ab-initio} $GW$+EDMFT calculations have shown that non-local correlation and screening effects are important for quantitatively accurate descriptions \cite{Petocchi2021,Petocchi2020a} and that they can drive the formation of electronic orders \cite{Ryee2020}. Also in the case of La$_3$Ni$_2$O$_7$, treating both the non-local and local physics in self-consistent $GW$+EDMFT has a significant effect on the electronic structure.
It triggers a charge-order instability which breaks the symmetry between the two layers with period-2 oscillations for the original primitive cell, at first sight suggestive of a staggered ordering 
(two sublattices with interchanged occupations, see SM for the pressure dependence).
If we allow the spins to rearrange freely, this period-2 solution 
shows a ferromagnetic (FM) ordering. The difference to the predicted AFM order found in EDMFT can be understood from the filling and the generic phase-diagram of the two-orbital Hubbard model \cite{Hoshino2016}, where AFM (FM) appears near 1/2 (3/4) filling.
However, the calculation for a supercell with two Ni dimers along the Ni--O--Ni bonds (see SM) again produces period-2 oscillations, which shows that the order is not of the staggered type, and that the oscillations rather indicate an instability to a more complex ordering pattern.

\begin{figure}
\begin{centering}
\includegraphics[width=\columnwidth]{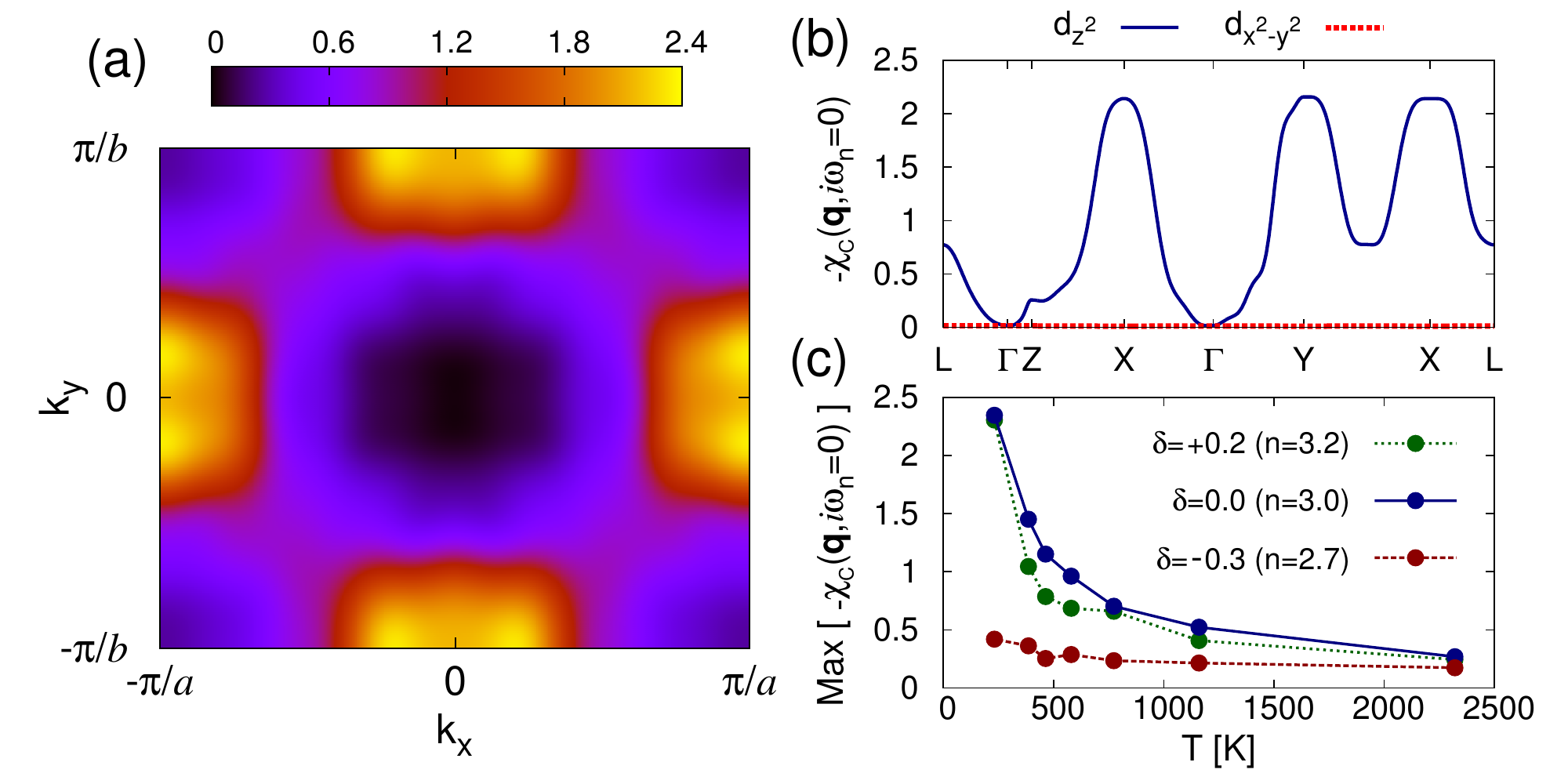}
\end{centering}
\vspace{-0.6cm}
\caption{
Charge susceptibility $-\chi_c({\bf q},i\omega_n=0)$ from $GW$+EDMFT with suppressed charge order at $\beta=50$~eV$^{-1}$ (a) in the ${\bf k}_{z}=0$ plane for $d_{x^2-y^2}$ and (b) along the indicated high-symmetry path.
(c) Temperature and doping dependence of the maximum of $-\chi_c({\bf q},i\omega_n=0)$.
\label{fig:ChiC}}
\end{figure}

To gain further insights into the ordering instability, and to get a better (beyond $GW$) description of the intra-bilayer correlations, we perform calculations for a two-site impurity cluster in a rotated bonding--anti-bonding basis (see SM).  By neglecting the off-diagonal hybridizations in this basis, we can suppress the charge-ordering tendencies, and measure the charge susceptibility $\chi_c({\bf q},i\omega_n)=[1-U_{\bf q}^\text{cRPA}(i\omega_n)\Pi_{\bf q}(i\omega_n)]^{-1}\Pi_{\bf q}(i\omega_n)$ in the symmetrized state.  
Here $\Pi_{\bf q}$ is the polarization from $GW$+EDMFT, and $\omega_n=2n\pi/\beta$ denotes Bosonic Matsubara frequencies.
In Fig.~\ref{fig:ChiC}(b), we show
$\chi_c({\bf q},i\omega_n=0)$ along the indicated high-symmetry path to demonstrate that a charge instability can be expected to occur in the $xy$-plane for the $d_{z^2}$-like orbital, while the $d_{x^2-y^2}$ orbital shows no ordering tendency.
Focusing on the $k_z=0$ plane in Fig.~\ref{fig:ChiC}(a), we find a multi-peak structure, with peaks centered around 
($\pm \frac{\pi}{a},\pm \delta$) and ($\pm \delta,\pm \frac{\pi}{b}$), with small $\delta$, 
indicating the formation of diagonal stripes. The susceptibility for this type of order grows as temperature is lowered [Fig.~\ref{fig:ChiC}(c)].
Along the stripes, we have a slow charge modulation with period $2\pi/\delta$ (in units of lattice spacing). 
Such an incommensurate charge density wave cannot be stabilized in simulations with a small unit cell. In particular, the ordering pattern is inconsistent with a simple staggering or with the doubled unit cell calculation that we have explicitly considered.

The reported results are robust against pressure changes in the $Fmmm$ phase, in the 20 GPa pressure range studied.
While charge ordering has so far not been addressed directly for the high-pressure phase (specific heat measurements provide indirect evidence for its disappearance) \cite{Wu2001,Liu2023,Zhang2023b,Sun2023}, it is interesting to note that charge and spin order has been experimentally detected in the low-pressure phase of La$_3$Ni$_2$O$_7$ \cite{Taniguchi1995,Wu2001,Liu2023,Zhang2023b}. 
Ref.~\cite{Taniguchi1995} discussed possible ordering tendencies and their dependence on the oxygen deficiency of the experimental structure. They also proposed diagonal stripes of the incommensurate type, and in particular they mentioned the possibility of a sine-wave modulated charge-ordering (charge density wave).
Furthermore, an early theoretical study of the low-pressure phase emphasized that the 1D character of parts of the DFT Fermi surface suggests an instability to (striped) charge order \cite{Seo1996}. 
An experimental confirmation of the presence or absence of charge stripes in the high-pressure phase would provide a check for the reliability of the $GW$+EDMFT predictions.

\begin{table}[b]
\caption{\label{Tab:doping} 
Doping dependence of the orbital occupations per spin from $GW$+EDMFT for the non-ordered and period-2 oscillating solution (filled site) at $\beta=50$ eV$^{-1}$. Negative (positive) values correspond to hole (electron) doping.
}
\setlength{\tabcolsep}{6pt} 
\renewcommand{\arraystretch}{1.5}
\centering
\begin{tabular}{|c||c c||c c|}
\hline
\multicolumn{1}{c}{} & \multicolumn{2}{c}{Suppressed order} & \multicolumn{2}{c}{Period-2 solution}  \\
\hline
Doping & $n_{x^2-y^2}$ & $n_{z^2}$ & $n_{x^2-y^2}$ & $n_{z^2}$  \\
\hline
\hline
$\phantom{-}$0.2 & 0.08 & 0.72 & 0.96 & 0.57 \\
$\phantom{-}$0.1 & 0.08 & 0.70  & 0.96 & 0.52 \\
$\phantom{-}$0.0 & 0.08 & 0.67  & 0.96 & 0.47 \\
$-0.1$ & 0.08 & 0.64 & 0.95 & 0.43 \\
$-0.2$ & 0.09 & 0.61  & 0.90  & 0.43  \\
$-0.3$ & 0.09 & 0.59 & 0.85  & 0.43  \\
\hline
\end{tabular}
\end{table}

The local spectral functions from the $GW$+EDMFT calculations are shown in Fig.~\ref{fig:spectra_ordering}(c).
Only the $d_{z^2}$ orbital has significant spectral weight 
close to the Fermi energy, both for the non-ordered and period-2 solutions. In the latter case the nearly half-filled $d_{z^2}$ orbital is found to be metallic, but close to a Mott transition, with a sharp quasi-particle peak split off from the doped upper Hubbard band, while in the 2/3-filled solution with suppressed order we find the familiar 3-peak structure of a moderately correlated metal (see also the self-energies in panel (d)). 
The dominance of the $d_{z^2}$ orbital distinguishes La$_3$Ni$_2$O$_7$ from the infinite-layered nickelates or the cuprates, where the single-band description involves the $d_{x^2-y^2}$ orbital. This should have profound implications for the pairing symmetry.

We finally return to the previously mentioned self-doping effect due to the La layers, which has been eliminated by the structural relaxation. It is interesting to ask where holes introduced by the presence of such a pocket would end up.
Self-consistent $GW$+EDMFT results for different doping levels (simulated by a shift of the chemical potential) are reported in Tab.~\ref{Tab:doping}. In the electron doped and hole doped regime, the doped charge carriers initially go into the $d_{z^2}$  orbital, while the almost empty or filled $d_{x^2-y^2}$ orbitals are little affected, especially in the case of suppressed charge order.
Interestingly, upon hole doping, the susceptibility is markedly reduced compared to the nominal situation with $n=3$ electrons  [Fig.~\ref{fig:ChiC}(c)]. 
Hence, if self-doping occurs in the real material under pressure, this would suppress the charge order and potentially favor the superconducting state.

{\it Conclusions} Our self-consistent $GW$+EDMFT results for La$_3$Ni$_2$O$_7$ under pressure produce a markedly different physical picture than that obtained from DFT and DMFT. In particular, the self-consistent treatment of local and nonlocal screening and correlation effects results in an electronic structure with the $d_{z^2}$ orbital
as the main active player. While the period-2 oscillating solutions appear to be an artifact of a too small unit cell, the charge susceptibility reveals that $GW$+EDMFT predicts a charge instability towards the formation of diagonal stripes in the $x$-$y$ plane, with the charge modulation in the $d_{z^2}$ orbitals.
In the broader context of high-temperature superconductivity, especially in cuprates, the role of fluctuating or static charge order has been much debated \cite{Keimer2015}.
If such a competing or co-existing order exists also in the present bilayer nickelate superconductor is a relevant question that merits further experimental investigations.
On the theory side, expanding the low-energy description to include La and O states will provide insights into the effects of the charge transfer from the O bands \cite{Lechermann2023} and the role of the self-doping $\Gamma$-pocket, which is found in the DFT calculations of the experimentally reported structure. This appears to be relevant since our calculations indicate a reduced charge ordering tendency upon hole doping.

{\it Acknowledgments} We acknowledge support from the Swiss National Science Foundation via NCCR Marvel and Grant No. 200021-196966.

\bibliography{main}%

\end{document}


\title{Supplemental Material: Correlated electronic structure of La$_3$Ni$_2$O$_7$ under pressure}
\author{Viktor Christiansson}
\affiliation{Department of Physics, University of Fribourg, 1700 Fribourg, Switzerland}
\author{Francesco Petocchi}
\affiliation{Department of Quantum Matter Physics, University of Geneva, 1211 Geneva 4, Switzerland}
\author{Philipp Werner}
\affiliation{Department of Physics, University of Fribourg, 1700 Fribourg, Switzerland}

\maketitle

\section{Computational details for the first principles calculations}

\begin{figure}[b]
\begin{centering}
\includegraphics[width=0.35\columnwidth]{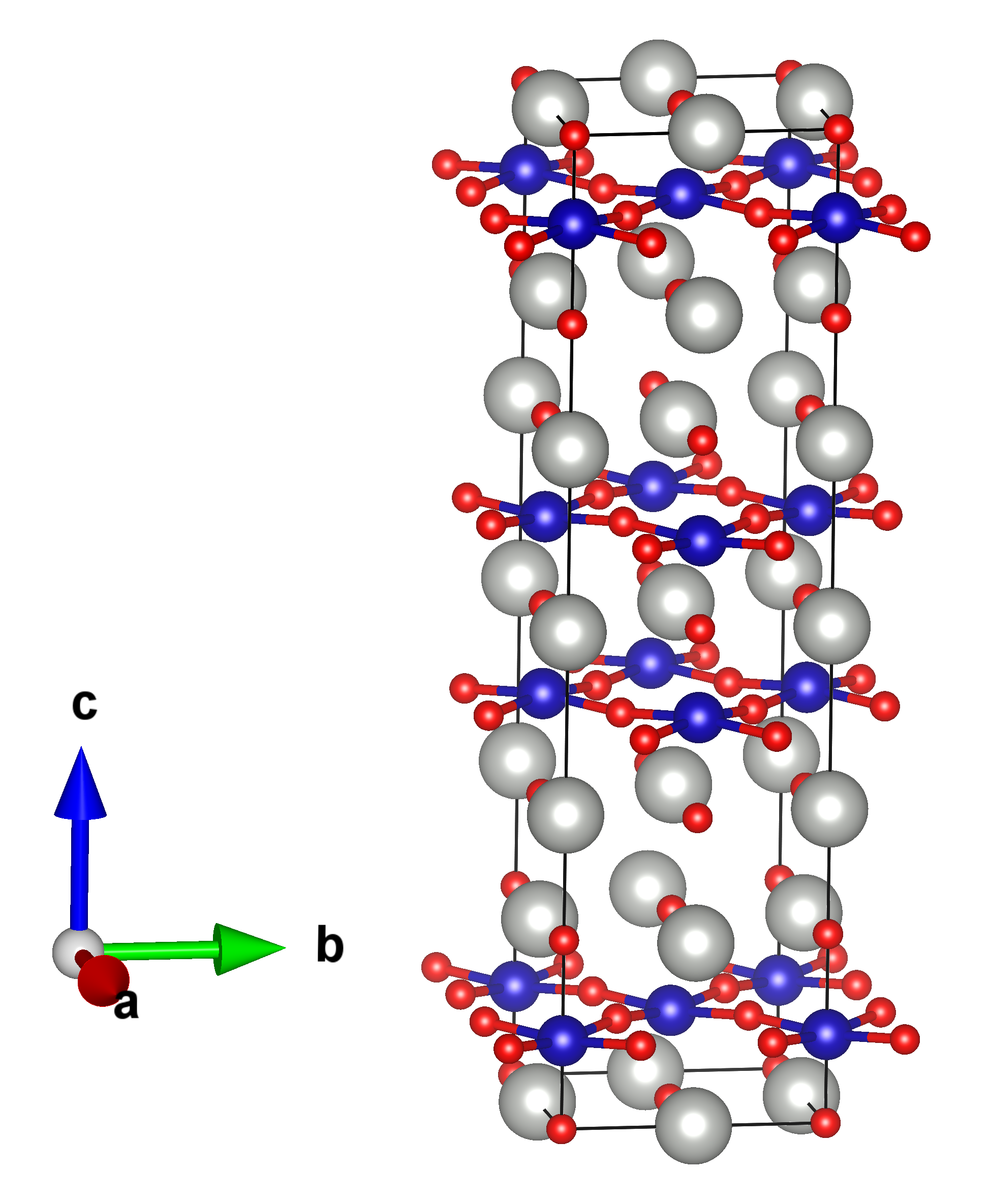}
\hspace*{1cm}
\includegraphics[width=0.50\columnwidth]{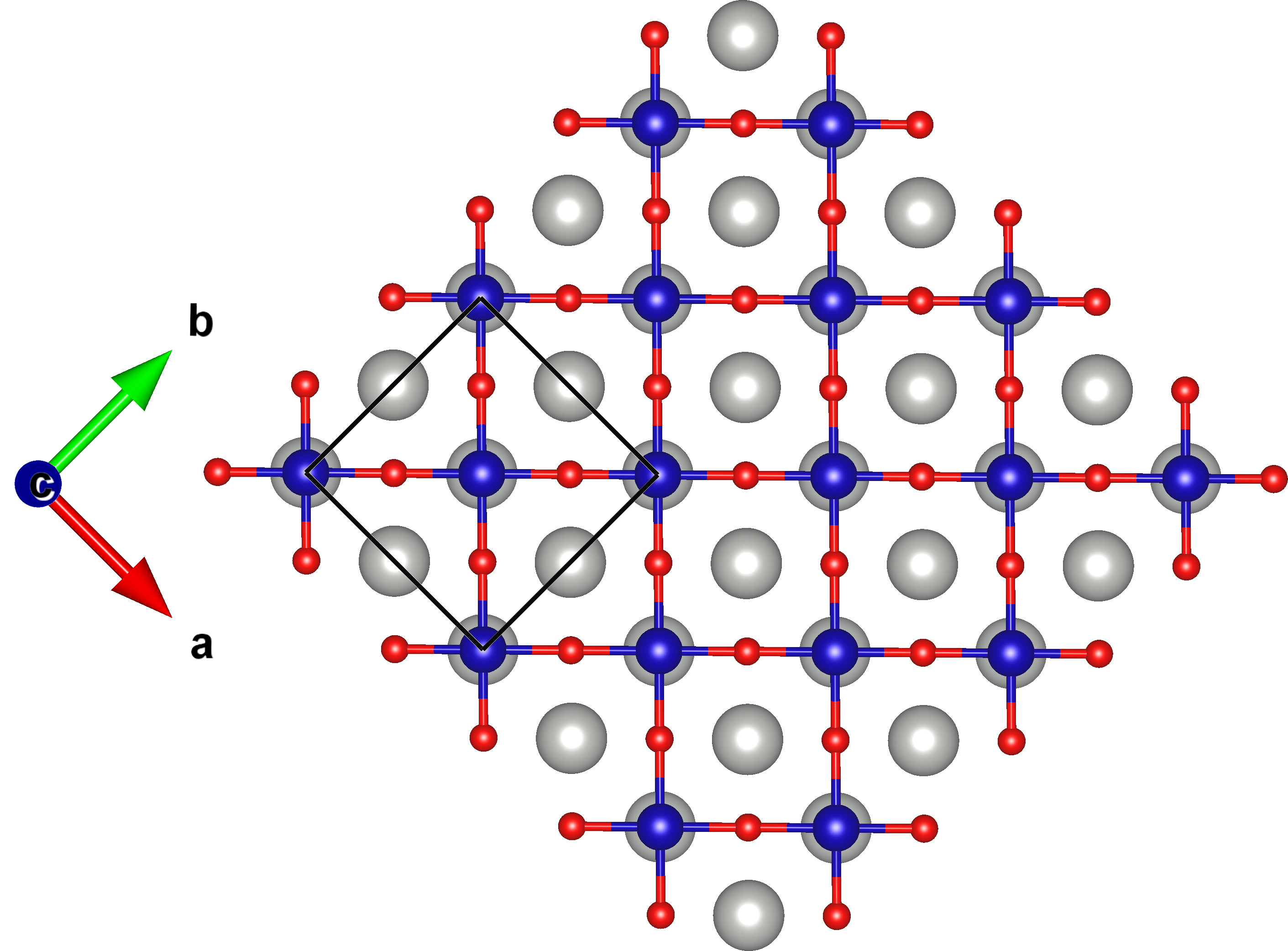}
\end{centering}
\caption{Left: Crystal structure of the high-pressure $Fmmm$ phase of La$_3$Ni$_2$O$_7$ drawn using the VESTA software \cite{VESTA}. The primitive cell contains a single formula unit. The bilayers of Ni (blue) and O (red), shown in the center, are separated by a La (gray) layer. The Ni--O bonds highlight the Ni squares within a single plane. Right: View of a Ni-O plane from above. The black square shows the conventional unit cell. We use a doubled unit cell with two dimers along the diagonal (here rotated) Ni--O--Ni bond for our supercell calculations.
\label{fig:structure}}
\end{figure}

In the density functional theory (DFT) calculations we use the experimentally reported \cite{Sun2023} high-pressure $Fmmm$ phase (space group 69) of bilayer La$_3$Ni$_2$O$_7$ containing a single formula unit, with one Ni atom in each of the two layers. The crystal structure is displayed in Fig.~\ref{fig:structure}. The pressure is determined by the experimentally reported lattice constants. 
The DFT calculations are performed using the full-potential all-electron code Fleur \cite{Fleurcode} with the generalized gradient approximation (GGA) \cite{Perdew1996} functional using a $16\times16\times16$ ${\bf k}$-grid.

As was mentioned in the main text, the DFT band structure is affected by the choice of atomic positions. The experimentally reported positions (for pressure $P=29.5$ GPa) \cite{Sun2023} lead to a self-doping pocket due to the La layer around the $\Gamma$-point. Upon relaxing the positions, the main effect is that this pocket is lifted above the Fermi energy by an approximately 1 eV shift.  Our relaxed positions for the three pressures we have considered are reported in Tab.~\ref{tab:parameters}. We note that the relaxed positions for $P=$29.5 GPa agree with those calculated in Ref.~ \cite{Sun2023}.

\begin{figure}[t]
\begin{centering}
\includegraphics[width=\columnwidth]{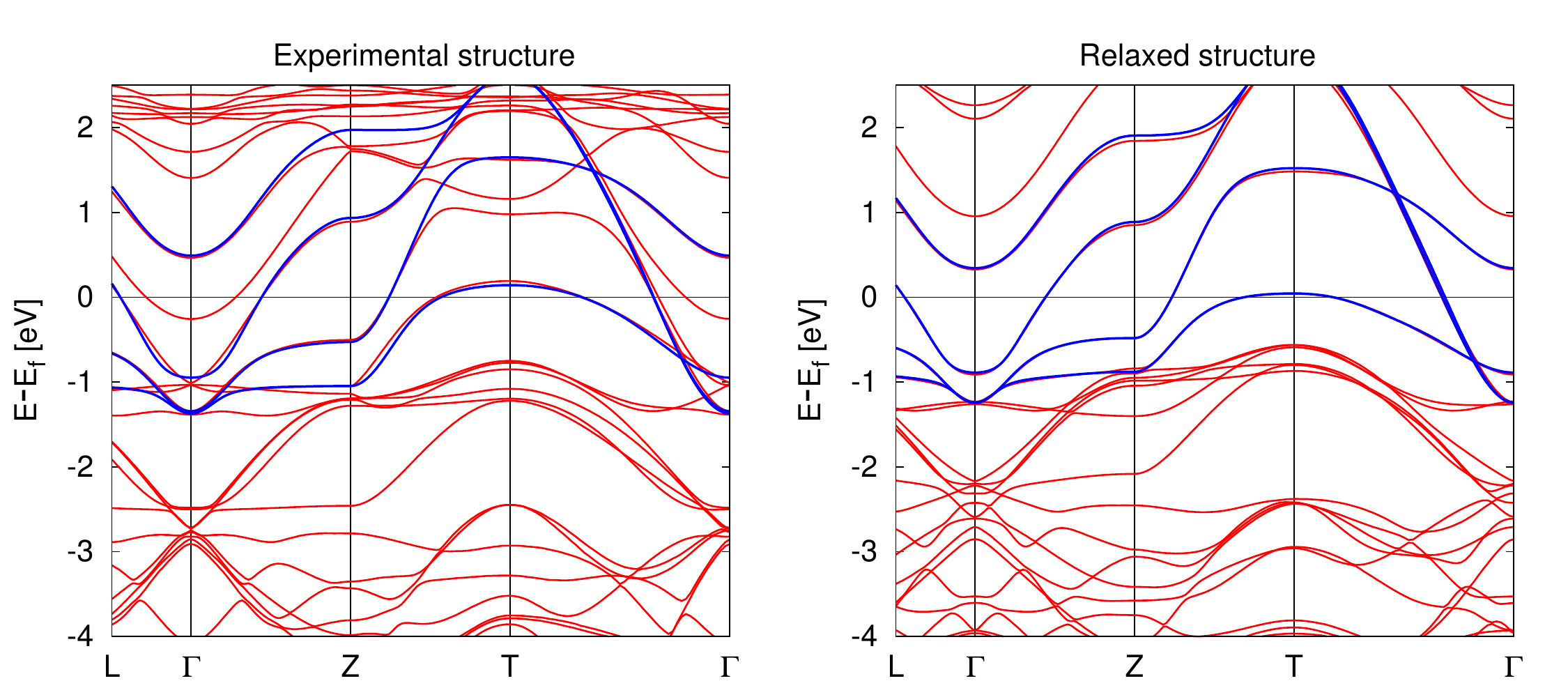}
\end{centering}
\caption{
Full bandstructure (red) and Wannier-derived bandstructure of the four-orbital model (blue) at $P=29.5$ GPa. The left panel is for the experimental structure. Here, the original bandstructure exhibits a hole pocket associated with La states near the $\Gamma$-point. The right panel is for the relaxed structure, where this band is lifted up by approximately 1~eV.
\label{fig:DFT_BS}}
\end{figure}

\newpage

\begin{table*}[h!]
\caption{Lattice parameters in \AA~and position parameters for Wyckoff positions in the relaxed structures of the high-pressure phase $Fmmm$ of La$_3$Ni$_2$O$_7$. The values in brackets for $P=29.5$ GPa are the experimentally reported positions in Ref.~\cite{Sun2023}. In the relaxation we have fixed the lattice constants to their experimental values at the three pressures $P=20.9$, 29.5, and 41.2 GPa considered in this work.
\label{tab:parameters}
}
\setlength{\tabcolsep}{15pt} 
\renewcommand{\arraystretch}{1.0} 
\centering
\begin{tabular}{|c|c||c|c|c||c|c|c|}
\hline
\multicolumn{4}{c}{$P=$20.9 ($a$=5.309  \AA, $b$=5.272  \AA, $c$=19.861  \AA)}    \\
\hline
\multicolumn{1}{c}{} & \multicolumn{1}{c}{x}  & \multicolumn{1}{c}{y}  & \multicolumn{1}{c}{z} \\
\hline
Ni & 0 & 0 & 0.097  \\
La(1) & 0 & 0 & 0.321  \\
La(2) & 0 & 0 & 0.5 \\
O(1) & 0.25 & 0.25 & 0.095  \\
O(2) & 0 & 0 & 0.204  \\
O(3) & 0 & 0 & 0 \\
\hline
\end{tabular}
\begin{tabular}{|c|c||c|c|c||c|c|c|}
\hline
\multicolumn{4}{c}{$P=$29.5 ($a$=5.289 \AA, $b$=5.218 \AA, $c$=19.734 \AA)}    \\
\hline
\multicolumn{1}{c}{} & \multicolumn{1}{c}{x}  & \multicolumn{1}{c}{y}  & \multicolumn{1}{c}{z} \\
\hline
Ni & 0 & 0 & 0.097 [0.089] \\
La(1) & 0 & 0 & 0.321 [0.316]  \\
La(2) & 0 & 0 & 0.5 \\
O(1) & 0.25 & 0.25 & 0.095 [0.096] \\
O(2) & 0 & 0 & 0.204 [0.190]  \\
O(3) & 0 & 0 & 0 \\
\hline
\end{tabular}
\begin{tabular}{|c|c||c|c|c||c|c|c|}
\hline
\multicolumn{4}{c}{$P=$41.2 ($a$=5.236 \AA, $b$=5.167 \AA, $c$=19.538 \AA)}    \\
\hline
\multicolumn{1}{c}{} & \multicolumn{1}{c}{x}  & \multicolumn{1}{c}{y}  & \multicolumn{1}{c}{z} \\
\hline
Ni & 0 & 0 & 0.098  \\
La(1) & 0 & 0 & 0.321  \\
La(2) & 0 & 0 & 0.5 \\
O(1) & 0.25 & 0.25 & 0.095 \\
O(2) & 0 & 0 & 0.204 \\
O(3) & 0 & 0 & 0 \\
\hline
\end{tabular}
\end{table*}

\newpage

To move beyond the DFT calculations, we use maximally localized Wannier functions  \cite{Marzari1997} from the Wannier90 library \cite{Mostofi2008} as basis functions to define our low-energy subspace. We use a minimal model of two Wannier functions of $d_{x^2-y^2}$ and $d_{z^2}$ character centered on each of the two Ni atoms (in different layers) in the primitive cell, defining a four-orbital low-energy model.
As shown in Fig.~\ref{fig:DFT_BS} the disentangled band structure (diagonalized Wannier Hamiltonian) agrees very well with the DFT band structure for the relaxed structure, which shows that this model is a suitable starting point for our many-body calculations.

The cRPA \cite{Aryasetiawan2004} and one-shot $G^0W^0$ \cite{Hedin1965} calculations, which we use to downfold the full band structure to the low-energy model, provide the effective bare cRPA interactions $U_{\bf k}^\textrm{cRPA}(\omega)$ and effective bare propagators $G^{0}_{\bf k}(\omega)$, respectively. These calculations are performed using the SPEX code \cite{Friedrich2010} on an $8\times8\times8$ momentum grid, and we consider unoccupied DFT bands up to $\sim 60$ eV when computing the polarization function and self-energy.

The self-consistent calculations in the low-energy subspace are performed using the multitier framework of Refs.\cite{Boehnke2016,Nilsson2017}, which effectively incorporates the high-energy bands through the frequency dependent $U_{\bf k}^\textrm{cRPA}(\omega)$ and $G^{0}_{\bf k}(\omega)$. We use the same $8\times8\times8$ ${\bf k}$-grid for the self-consistency and for the downfolding. Since these calculations are performed at nonzero temperature, with frequency dependent self-energies, the dependence on the momentum resolution should in general be less important.

%
\section{Summary of the multitier $GW$+EDMFT method}

To calculate the electronic structure within the multitier $GW$+EDMFT formalism \cite{Boehnke2016,Nilsson2017}, we first run a DFT calculation which defines the (mean-field) starting point. The corresponding Kohn-Sham eigenvalues $\varepsilon_{n{\bf k}}$ and eigenfunctions $\varphi_{n{\bf k}}$ allow to define the low-energy model subspace (the four band model of the main text) using maximally-localized Wannier functions \cite{Marzari1997,Mostofi2008}. $G^0W^0$ \cite{Hedin1965} and cRPA \cite{Aryasetiawan2004} downfolding calculations yield the bare propagator $G^0_{\bf k}(\omega)$ and effective bare interaction $U_{\bf q}(\omega)$ of this low-energy subspace. 

In the low-energy space, we compute the self-energy and polarization by combining the $GW$ results with more accurate local contributions from an EDMFT impurity calculation, using a well-defined double counting.
The 
local correlated problem (corresponding to the two Ni sites)
is solved using the EDMFT self-consistency conditions, $G_\textrm{imp}=G_\textrm{loc}$ and $W_\textrm{imp}=W_\textrm{loc}$, where the subscripts refer to the impurity and local  lattice quantities, respectively. The contributions from the rest space (all states outside of the low-energy space) are taken into account via the initial $G^0_{\bf k}(\omega)$ and $U_{\bf q}(\omega)$ on the one-shot $GW$ and RPA ($\Pi=G^0G^0$) level.

The explicit form of the interacting lattice Green's function $G_{\bf k}$ and screened interaction $W_{\bf k}$ is 
\begin{align}
G^{-1}_{{\bf k}}=&\,  i\omega_n + \mu -\varepsilon_{{\bf k}}^{\textrm{DFT}} + V^\text{XC}_{{\bf k}}  - \left(\Sigma_{{\bf k}}^{G^0W^0} - \Sigma_{{\bf k}}^{G^0W^0}\big|_{I} \right) \nonumber \\
& - \left( \Sigma_{{\bf k}}^{\textrm{sc}GW}\big|_{I} - \Sigma_{\mathrm{loc}}^{\textrm{sc}GW}\big|_{C} \right)-\Sigma_{\mathrm{loc}}^\textrm{EDMFT}\big|_{C},
\label{Eq:Gint}
\end{align}
\begin{align}
W^{-1}_{{\bf k}}=& v_{{\bf k}}^{-1}- \left(\Pi_{{\bf k}}^{G^0G^0} - \Pi_{{\bf k}}^{G^0G^0}\big|_{I} \right) \nonumber \\
& - \left( \Pi_{{\bf k}}^{GG}\big|_{I} - \Pi_{\mathrm{loc}}^{GG}\big|_{C} \right  )-\Pi_{\mathrm{loc}}^\textrm{EDMFT}\big|_{C}.
\label{Eq:Wint}
\end{align}\\
%
Here $I$ refers to the intermediate space (in this case the four orbital model) and $C$ to the local correlated space (two separate impurity problems for each of the Ni sites). 
Note the well-defined double counting between the different spaces \cite{Nilsson2017}.
In the calculation of the lattice quantities, the full frequency- and momentum-dependence as well as all the Green's function and interaction matrix elements are taken into account. 
The two Ni sites are thus coupled in the fermionic and bosonic self-consistency equations,
in which the impurity quantities can be considered as providing local site-dependent self-energy and polarization vertex corrections beyond $GW$.

In the impurity model calculations (one for the Ni in each layer), we use at each iteration in the self-consistency updated fermionic (hybridization function) and bosonic (retarded interaction) Weiss fields derived for the corresponding sites,
\begin{align}
	\mathcal{G}_i=\left(\Sigma_i^\mathrm{loc}+(G_i^\mathrm{loc})^{-1}\right)^{-1} & \qquad i\in\left\{ \mathrm{Ni}_1,\mathrm{Ni}_2\right\}  , \label{eq:GWeiss}\\
\mathcal{U}_i=W_i^\mathrm{loc}\left(1+\Pi_i^\mathrm{loc}W_i^\mathrm{loc}\right)^{-1} & \qquad i\in\left\{ \mathrm{Ni}_1,\mathrm{Ni}_2\right\}.    \label{eq:WWeiss} 
\end{align}
Here the $G_i^\text{loc}$, $\Pi_i^\mathrm{loc}$, and $W_i^\text{loc}$ are computed in the previous iteration as the $2\times2$ submatrices of the correpsonding local interacting lattice quantities. 
The resulting impurity problem in the EDMFT calculations is then solved using a continuous-time Monte Carlo solver \cite{Werner2006,Hafermann2013} capable of treating models with dynamically screened interactions \cite{Werner2010}.

\section{Calculations in the rotated basis}

To treat inter-layer nonlocal correlations within EDMFT, we solve a two-site impurity problem in a rotated bonding--anti-bonding basis.
Here we use the rotation which diagonalizes the local Hamiltonian
\begin{equation}
\frac{1}{N_{\bf k}}\sum_{\bf k} H_\textrm{DFT}({\bf k})=H({\bf R}=0)=
\begin{pmatrix}
  0.788 &  0.0 & 0.011 &  0.0\\ 
  0.0&  0.413 & 0.0 & -0.653 \\
  0.011 & 0.0 & 0.788 & 0.0 \\
  0.0 &-0.653 & 0.0 &  0.413
\end{pmatrix}.
\end{equation}
As long as the two sites remain equivalent (no symmetry breaking) this rotation results in a diagonal hybridization function, which is required by our impurity solver.
The rotation $E_\textrm{loc} =\theta^\dagger H_\textrm{loc} \theta$ with
%
%
\begin{equation}
\theta\approx
\begin{pmatrix}
0 & -\frac{1}{\sqrt{2}} & -\frac{1}{\sqrt{2}} & 0 \\
\frac{1}{\sqrt{2}}& 0 & 0 & \frac{1}{\sqrt{2}} \\
0 & \frac{1}{\sqrt{2}} & -\frac{1}{\sqrt{2}} & 0 \\
\frac{1}{\sqrt{2}} & 0 & 0 & \frac{1}{\sqrt{2}} \\
\end{pmatrix}
\end{equation}
results in four molecular orbitals, with a $\sim 1.3$ eV separation between the $d_{z^2}$ bonding and anti-bonding orbitals: 
\begin{equation}
E^\textrm{rotated}_\textrm{loc}=\begin{pmatrix}
-0.240 \\
0.776 \\
0.799 \\
1.066
\end{pmatrix}.
\end{equation}


The basis transformation generates nontrivial interactions between these molecular orbitals, like pair-hopping and spin-flip terms, but in the EDMFT calculations, we only treat the density-density contributions.  

In the rotated basis, a charge ordered solution would feature large off-diagonal hybridizations.
By dropping these off-diagonal terms, the rotated-basis calculation allows us to stabilize a non-ordered solution (using mixing) and to measure the charge susceptibility as discussed in the main text. 
Without mixing, 
a spontaneous symmetry breaking eventually occurs in the lattice quantities, resulting in orbital occupations consistent with those found in the calculation with two separate impurity models.
Since we do not treat the off-diagonal hybridizations in the impurity calculation, this can however not be considered a fully converged solution.

\section*{\quad Pressure dependence of the interaction parameters}

In the $GW$+EDMFT calculations reported in the main text, the full non-local and frequency-dependent interaction tensor
\begin{align}\label{eq:U_wannier}
U_{n_1n_2n_3n_4}(\omega,{\bf R}) &= \int \int \mathrm{d}{\bf r}\mathrm{d}{\bf r}' \varphi_{n_1{\bf 0}}^*({\bf r})\varphi_{n_2{\bf 0}}({\bf r})U({\bf r},{\bf r}',\omega)\varphi^*_{n_3{\bf R}}({\bf r}')\varphi_{n_4{\bf R}}({\bf r}') \nonumber \\
&= \frac{1}{N_{q}}\sum_{\bf{q}} e^{-i \bf{q}\bf{R}} \langle \Psi_{n_1n_2}({\bf q},r)| U(\omega)| \Psi_{n_3n_4}({\bf q},r')\rangle  
\end{align} 
is used for the bare ($U$) and screened ($W$) interactions. Here the localized single-particle basis $\varphi_{n{\bf R}}(r)$ corresponds to maximally localized Wannier functions, and $\Psi_{n_1n_2}({\bf q},r)$ is the reduced product basis used in Ref.~\cite{Nilsson2017}.

In the main text we reported the on-site interaction parameters at $P=29.5$~GPa calculated within the constrained random-phase approximation (cRPA) \cite{Aryasetiawan2004}. For completeness, we report here for the three pressures considered the static local values of all the density-density terms $U_{nn,mm}(\omega=0,{\bf R}=0)$ in eV, including the inter-site (intra-dimer) interactions along the $z$-direction between the two sites in different layers. The superscript $i=1$ is for the upper layer and $2$ for the lower one, with the diagonal terms corresponding to intra-orbital and the off-diagonal terms to inter-orbital interactions (thus both intra- and inter-site terms within the dimer are presented):

\begin{equation*}
U^{P=20.9 \textrm{ GPa}}_{\mathrm{cRPA}}(0){=}\begin{pmatrix}
 d_{x^2-y^2}^{1} & \quad d_{z^2}^1 &  d_{x^2-y^2}^{2} & \quad d_{z^2}^2  \\
\hline
3.74 & 2.36 & 0.50 & 0.60   \\
2.36  & 3.54 & 0.60 & 0.80   \\
0.50 & 0.60 & 3.74 & 2.36   \\
0.60 & 0.80 & 2.36 & 3.54  
\end{pmatrix},\quad
\label{eqn:UcRPA_P20} 
\end{equation*}

\begin{equation*}
U^{P=29.5 \textrm{ GPa}}_{\mathrm{cRPA}}(0){=}\begin{pmatrix}
 d_{x^2-y^2}^{1} & \quad d_{z^2}^1 &  d_{x^2-y^2}^{2} & \quad d_{z^2}^2  \\
\hline
3.79 & 2.39 & 0.51 & 0.61    \\
2.39 & 3.58 & 0.61 & 0.81 \\
0.51 & 0.61 & 3.79 & 2.39 \\
0.61 & 0.81 & 2.39 & 3.58
\end{pmatrix},\quad
\label{eqn:UcRPA_P29} 
\end{equation*}

\begin{equation*}
U^{P=41.2 \textrm{ GPa}}_{\mathrm{cRPA}}(0){=}\begin{pmatrix}
 d_{x^2-y^2}^{1} & \quad d_{z^2}^1 &  d_{x^2-y^2}^{2} & \quad d_{z^2}^2  \\
\hline
3.79 & 2.38 & 0.51 & 0.61  \\
2.38  & 3.56 & 0.61 & 0.81  \\
0.51  & 0.61 & 3.79 & 2.38  \\
0.61  & 0.81 & 2.38 & 3.56  
\end{pmatrix}.\quad
\label{eqn:UcRPA_P41} 
\end{equation*}

We see that the interaction parameters for the two Ni atoms are identical, which reflects the fact that the localized basis states are locally equivalent for the two sites. We further find only very minor pressure induced differences in the calculated interaction parameters in the $\sim 20$ GPa pressure range studied. Also the on-site effective Hund's coupling ($J=U_{nm,mn}(\omega=0,{\bf R}=0)$) is hardly affected by the applied pressure: $J^{20.9 \textrm{ GPa}}_{\mathrm{cRPA}}=0.60$ eV, $J^{29.5 \textrm{ GPa}}_{\mathrm{cRPA}}=0.61$ eV, and $J^{41.2 \textrm{ GPa}}_{\mathrm{cRPA}}=0.61$ eV.

\section*{Non-local interaction $U({\bf R})$ from \lowercase{c}RPA}

Within the cRPA formalism \cite{Aryasetiawan2004}, the polarization function $\Pi$ is split into two parts, one for the low-energy model contributions to the screening processes ($\Pi_d$) and one for the ``rest" space ($\Pi_r$), the latter encompassing both screening channels between states outside of the model subspace, and between the model and the outside space.
The effective bare interaction $U$ is then calculated as
\begin{equation}\label{eq:UcRPA_Pr}
U(\omega)=[1-v\Pi_r(\omega)]^{-1} v,
\end{equation}
where $v$ is the bare Coulomb interaction. Since the low-energy metallic screening originating from the model subspace is eliminated in $\Pi_r=\Pi-\Pi_d$, the calculated effective bare interaction for the low-energy model typically displays a long-ranged Coulomb tail as shown in Fig.~2 of the main text. In Table~\ref{tab:nonlocalU} we report the interaction parameters for the $d_{x^2-y^2}$- and $d_{z^2}$-like orbitals both in the Ni-planes and perpendicular to them at several lattice spacings. As discussed in the main text, the nearest-neighbor interactions are relatively large, indicating that non-local screening and correlation effects may play an important role in this system.

\begin{table*}
\caption{Calculated static values of the non-local interaction $U(|{\bf R}|)$ at different lattice spacings $|{\bf R}|$ in \AA~in the plane ($\parallel$) and in the $z$-direction (perpendicular to the planes, $\perp$) for the $d_{x^2-y^2}$- and $d_{z^2}$-like orbitals.
\label{tab:nonlocalU}
}
\setlength{\tabcolsep}{8pt} 
\renewcommand{\arraystretch}{1.0} 
\centering
\begin{tabular}{|c||c c|| c||c  c|}
\hline
$|{\bf R}_\parallel|$ [\AA] & $U^{x^2-y^2}_{\parallel}$ [eV] & $U^{z^2}_{\parallel}$ [eV] & $|{\bf R}_\perp|$ [\AA] &  $U^{x^2-y^2}_{\perp}$ [eV] & $U^{z^2}_{\perp}$ [eV] \\
\hline
\hline
 0 &     3.79    & 3.58     & 0 & 3.79 & 3.58  \\
 3.7    &   0.76 & 0.59  & 3.8 & 0.51 & 0.81 \\
 5.2       &  0.48 & 0.42& 15.9 & 0.13 & 0.14   \\
 7.4       &  0.33 & 0.31 & 19.7 & 0.08 & 0.08 \\
 8.4      &    0.29 & 0.28& 23.5 & 0.04 & 0.04 \\
 10.4        &  0.24 & 0.23& 35.7 & 0.00 & 0.00 \\  
 11.14   &   0.23 & 0.22 & && \\ 
 11.8 & 0.22 & 0.21 & &&\\
 13.3 & 0.20 & 0.19& && \\
 15.7 & 0.18 & 0.18& && \\
     \hline
\end{tabular}
\end{table*}

\section{Local occupation and spin statistics}
\begin{figure}[b]
\begin{centering}
\includegraphics[width=0.65\columnwidth]{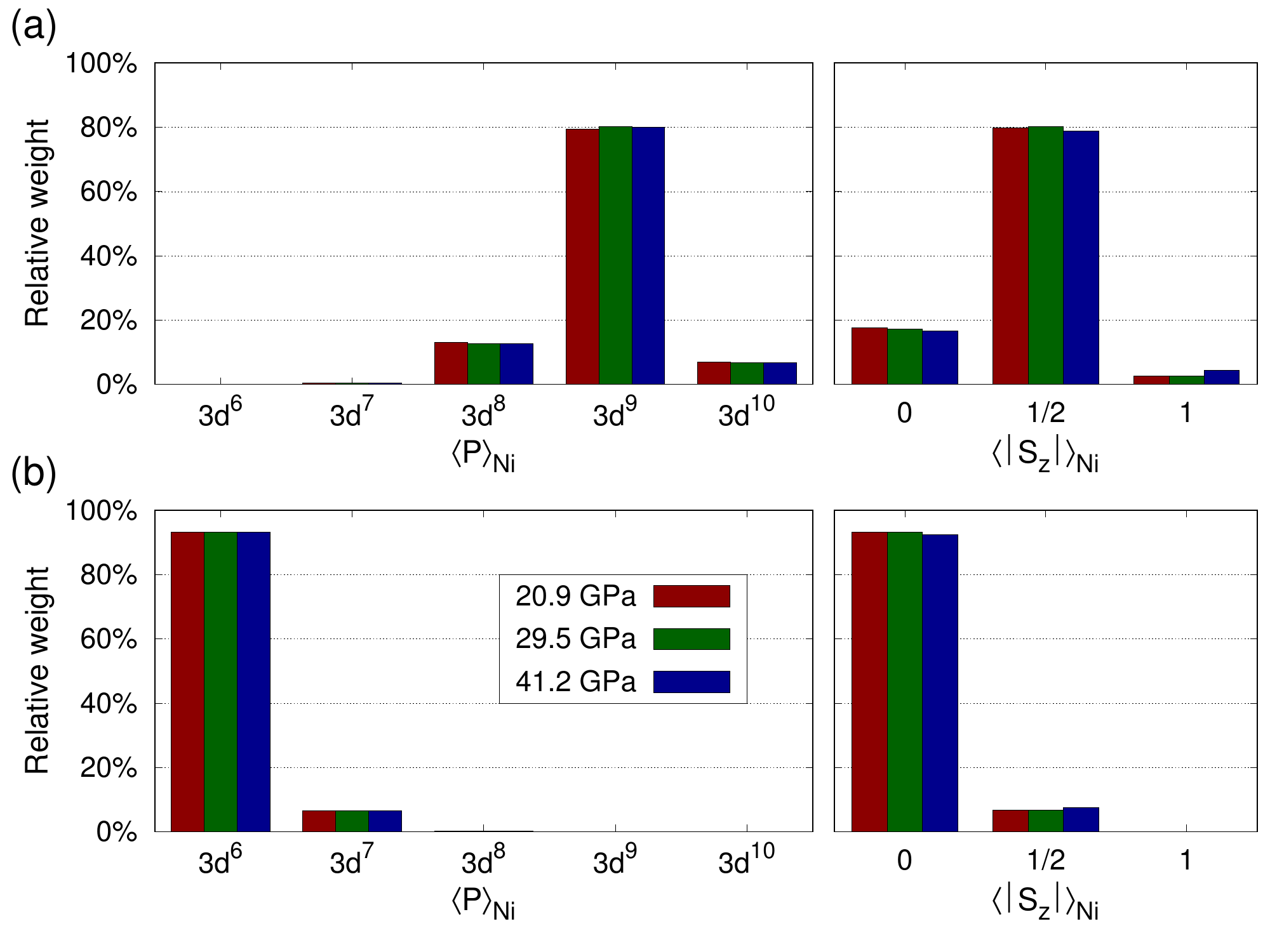}
\end{centering}
\vspace*{-0.6cm}
\caption{$GW$+EDMFT results at $\beta=10$ for the occupation and spin statistics on the
(a) filled site and (b) empty site. The different colors correspond to the pressures indicated in the label.  
The results for the other sublattice are equivalent. 
\label{fig:histograms}}
\end{figure}

In this section, we show the histograms of the local charge and spin states for the two atomic sites and for the three pressures listed in the label of Fig~\ref{fig:histograms}. These are results for the primitive cell calculations with period-of-two oscillations. The statistics thus reflects an ordered state induced by the constraints of a small unit cell, which is different from the stripe order predicted by the susceptibility analysis, as discussed in the main text.

The histograms show that on the considered sublattice, the Ni$_{1}$ atom is close to a $3d^{9}$ configuration, with only small fluctuations to higher and lower charge states. The Ni$_{2}$ atom instead is almost completely empty (corresponding to a dominant $3d^{6}$ configuration). Here Ni$_{i}$ with $i=1$, 2 denotes the two now inequivalent atoms in the two layers. 
These results agree with the expectations based on the orbital occupations in Tab.~\ref{Tab:Occupations}. The distribution of charge states remains robust throughout the high-pressure phase, and the occupations do not show any significant temperature dependence down to $\beta=50$~eV$^{-1}$. 
The spin statistics, together with the occupations in Tab.~\ref{Tab:Occupations} reveals a single active Ni $3d_{z^2}$ orbital (with occupation $\sim 0.5$ and spin moment $|S_z| \sim 0.5$) per unit cell in the period-of-two oscillating charge ordered state. 

\section{Supercell calculations}

In our supercell calculations we use a doubled unit cell containing two formula units which coincides with the primitive cell of the low-pressure $Amam$ phase. We have fixed the lattice parameters and relaxed atomic positions to the ones used above for the primitive cell at $P=29.5$ GPa. The two Ni atoms per layer are with this choice positioned along the same Ni--O--Ni bonds (forming two dimers in series), see Fig.~\ref{fig:structure}. 

The low-energy subspace for the supercell calculations is constructed as for the primitive cell calculations with $d_{x^2-y^2}$- and $d_{z^2}$-like Wannier functions on each of the four Ni sites, in total defining an eight-orbital model.
In the EDMFT part, we then solve four separate impurity problems (one for each Ni site), which are connected in the self-consistency loop through the (nonlocal) lattice quantities.
We use a $5\times5\times5$ ${\bf k}$-grid and unoccupied DFT bands up to $\sim 60$ eV for both the polarization function and self-energy in the downfolding using cRPA and $G^0W^0$.
The same momentum grid is used for the self-consistent $GW$+EDMFT calculations.

\begin{figure}[t]
\begin{centering}
\includegraphics[width=0.75\columnwidth]{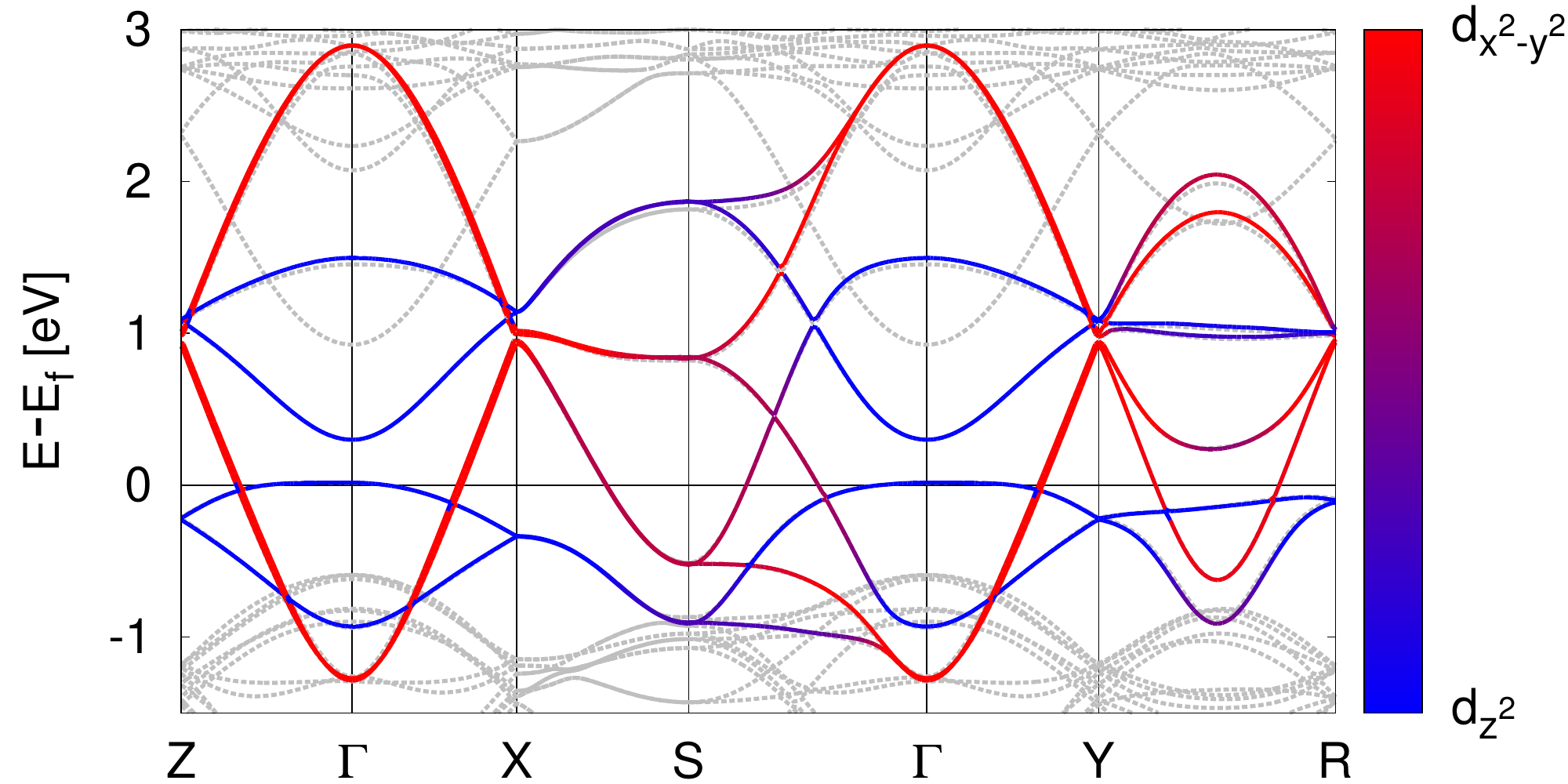}
\end{centering}
\caption{
The model band structure (solid lines) for the eight-orbital supercell calculations plotted on top of the DFT band structure (gray dashed lines) at $P=29.5$ GPa. The color-bar shows the orbital character of the model bands.
\label{fig:supercell_BS}}
\end{figure}

The interaction parameters for the supercell calculated within cRPA agrees very well with the ones reported above for the primitive cell. Although the spreads of the localized Wannier functions differ slightly compared to the primitive cell calculations, the static local (${\bf R}=0$) interaction parameters are converged to within approximately 0.05 eV (in this case the interactions between orbitals centered on different Ni-dimers correspond to the nearest-neighbor interactions in the plane reported for the primitive cell). We show here the density-density interaction elements for one of the two equivalent dimers in the supercell:
\begin{equation*}
U^\textrm{Supercell}_{\mathrm{cRPA}}(0){=}\begin{pmatrix}
 d_{x^2-y^2}^{1} & \quad d_{z^2}^1 &  d_{x^2-y^2}^{2} & \quad d_{z^2}^2  \\
\hline
3.78 & 2.40 & 0.51 & 0.61   \\
2.40  & 3.59 & 0.61 & 0.81   \\
0.51 & 0.61 & 3.78 & 2.40   \\
0.61 & 0.81 & 2.40 & 3.59
\end{pmatrix}.
\label{eqn:UcRPA_supercell} 
\end{equation*}
The Hund's coupling between the $d_{x^2-y^2}$ and $d_{z^2}$ orbitals on a single site is again $J=0.61$ eV, in agreement with the primitive cell. We also remark that the inter-dimer interactions $U^{x^2-y^2}_\textrm{n.n.}=0.75$ and $U^{z^2}_\textrm{n.n.}=0.59$ eV agree with the (nearest-neighbor in the plane) interactions reported in Tab.~\ref{tab:nonlocalU}.

\begin{table*}[ht]
\caption{\label{Tab:Occupations} 
Orbital occupation per spin $n^i$ at inverse temperature $\beta=50$ eV$^{-1}$ ($\sim 230$ K) for the Ni$_i$ $3d_{x^2-y^2}$- and $3d_{z^2}$-like orbitals in the two layers $i$ of the primitive unit cell.
The total filling is $n=3$. Sublattice A and B refer to the electronic orderings for the period-2 oscillating solution discussed in the main text.
}

\setlength{\tabcolsep}{6pt} 
\renewcommand{\arraystretch}{1.15} 
\centering
\begin{tabular}{|c|c|c||c c||c c|}
\hline
Method & $P$ [GPa] &  & $n^{1}_{x^2-y^2}$ & $n^{1}_{z^2}$ & $n^{2}_{x^2-y^2}$ & $n^{2}_{z^2}$ \\
\hline
\hline

& 20.9 & & 0.32 & 0.43  & 0.32  & 0.43   \\
DFT   & 29.5 & & 0.31 & 0.44 & 0.31 & 0.44  \\
& 41.2 & & 0.31 & 0.44 & 0.31 & 0.44  \\
\hline
\hline

& 20.9 & & 0.23 & 0.52  & 0.23 & 0.52  \\
sc$GW$ & 29.5 & & 0.23 & 0.52 & 0.23 & 0.52  \\
&  41.2 & & 0.24 & 0.51 & 0.24 & 0.51  \\
\hline
\hline

& 20.9 & & 0.33 & 0.42 & 0.33 & 0.42  \\
DFT+DMFT $U^\textrm{cRPA}(\omega=0)$ & 29.5 & & 0.32 & 0.43 & 0.32 & 0.43  \\
& 41.2 & & 0.32 & 0.43 & 0.43  & 0.43   \\
\hline
\hline

& 20.9 & & 0.37  & 0.38 & 0.37 & 0.38  \\
DFT+DMFT $U^\textrm{cRPA}(\omega)$ & 29.5 &  & 0.37 & 0.38 & 0.37 & 0.38  \\
& 41.2 & & 0.37 & 0.38 & 0.37  & 0.38   \\
\hline
\hline

 & 20.9 &  &0.50 & 0.50 & 0.36 & 0.14  \\
EDMFT & 29.5 & &  0.50  & 0.50 & 0.34 & 0.16  \\
 & 41.2 & & 0.50 & 0.50 & 0.33 & 0.17  \\
\hline
\hline

$GW$+EDMFT & 20.9 & Sublattice A & 0.96  & 0.47 & 0.04 & 0.03  \\
  & 20.9 & Sublattice B & 0.04  & 0.03  & 0.96 & 0.47  \\
  \hline
$GW$+EDMFT & 29.5 & Sublattice A & 0.96 & 0.47 & 0.04  & 0.03   \\
 & 29.5 & Sublattice B & 0.04  & 0.03 & 0.96  & 0.47  \\
 \hline
$GW$+EDMFT  & 41.2 & Sublattice A & 0.96 & 0.48 & 0.03 & 0.03  \\
  & 41.2 & Sublattice B & 0.04  & 0.03  & 0.47 & 0.96  \\
\hline

\end{tabular}
\end{table*}

\section*{Pressure dependence of the orbital occupations}

As reported in the main text, our results are very robust throughout the high-pressure $Fmmm$ phase, and we show the full pressure dependence of our results in Tab.~\ref{Tab:Occupations}. We have checked these occupations over $\sim10$ converged iterations and they show deviations $<0.01$ electrons/spin, which 
is within the ``error bar" of our calculations. 

Since the effective bare interaction displays a significant frequency dependence, as shown in Fig.~2 of the main text, we consider different static $U$ values in the DFT+DMFT $U(\omega=0)$ calculations. We vary $U$ in the physically reasonable range from 3.2 eV to 5.7 eV. Since the screening of the Hund's coupling $J$ is less efficient, as demonstrated in the inset of Fig.~2, we keep this parameter fixed. The results remain consistent with the static cRPA calculation, see Tab.~\ref{Tab:Occupations_staticU}, showing that the DFT+DMFT picture is not very sensitive to the details of the interaction parameters. 
However, as discussed in the main text, the inclusion of nonlocal screening processes does have a significant effect. 
We note that in principle it is possible to define an effective static interaction parameter which mimics the frequency-dependent $U(\omega)$. This procedure however also requires a corresponding renormalization of the hopping parameters \cite{Casula2012}, which is difficult to implement for multi-orbital systems. 

\begin{table*}[ht]
\caption{\label{Tab:Occupations_staticU} 
Orbital occupations per spin $n^i$ at inverse temperature $\beta=50$ eV$^{-1}$ ($\sim 230$ K) for DFT+DFMT with varying $U(\omega=0)$ and fixed Hund's coupling $J=0.6$ eV at $P=29.5$ GPa. The total filling is $n=3$, and the Ni$_i$ $3d_{x^2-y^2}$- and $3d_{z^2}$-like orbitals are reported for the two layers $i$ in the primitive unit cell.
}

\setlength{\tabcolsep}{6pt} 
\renewcommand{\arraystretch}{1.15} 
\centering
\begin{tabular}{|c||c c||c c|}
\hline
$U$ [eV] & $n^{1}_{x^2-y^2}$ & $n^{1}_{z^2}$ & $n^{2}_{x^2-y^2}$ & $n^{2}_{z^2}$ \\
\hline
\hline
$3.2$ & 0.33 & 0.42 & 0.44 & 0.42   \\

$3.7$ & 0.33 & 0.42 & 0.33 & 0.42  \\

$U^\textrm{cRPA}(\omega=0)$ &0.32 & 0.43 & 0.32 & 0.43  \\

$4.2$ & 0.32 & 0.43 & 0.32 & 0.43   \\

$4.7$ & 0.32 & 0.43  & 0.32 & 0.43   \\

$5.2$ & 0.31 & 0.44 & 0.31 & 0.44    \\

$5.7$ & 0.31 & 0.44  & 0.31  & 0.44   \\
\hline
\end{tabular}
\end{table*}